\documentclass[conference, comsoc]{IEEEtran}
\IEEEoverridecommandlockouts
\usepackage[left=0.625in,right=0.625in,top=0.75in,bottom=1in,columnsep=0.21in]{geometry}

\usepackage{balance}
\usepackage{caption}
\usepackage{subcaption}
\usepackage{multirow}
\usepackage{array}
\usepackage{balance}
\usepackage{cite}
\usepackage{amsmath,amssymb,amsfonts}
\usepackage{graphicx}
\usepackage{booktabs}
\usepackage{textcomp}
\usepackage{bm}
\usepackage{nicefrac}
\usepackage{mathtools}
\usepackage{comment}
\usepackage{pgfplots}
\usepackage{pgfkeys,pgfmath,pgfcore}
\pgfplotsset{compat=1.18}
\pgfkeys{/pgf/number format/.cd,fixed,precision=2}

\usepackage{multirow}
\usepackage{xcolor} 
\usepackage{colortbl}
\usepackage{threeparttable}
\usepackage{dblfloatfix} 
\usepackage{acro}
\definecolor{verylightgray}{rgb}{0.9,0.9,0.9}
\usepackage{tikz}
\usetikzlibrary{arrows.meta,bending,calc}
\usepackage{algorithmic}
\usepackage{algorithm}
\usepackage{acro}

\DeclareAcronym{sdr}{
    short = SDRs,
    long = Software-Defined Radios
}

\DeclareAcronym{ema}{
    short = EMA,
    long = Exponential Moving Average
}

\DeclareAcronym{jepa}{
  short = JEPA,
  long = Joint-Embedding Predictive Architecture
}

\DeclareAcronym{mae}{
  short = MAE,
  long = Masked Autoencoder
}

\DeclareAcronym{simclr}{
  short = SimCLR,
  long = Simple Contrastive Learning of Representations
}

\DeclareAcronym{vit}{
    short = ViT,
    long = Vision Transformer
}

\begin{document}

\title{Representation Diversity for Wireless Foundation Models}

\title{ Wireless Foundation Model Scaling \\ via Representation Diversity and\\ Multi-Branch Architectures}

\title{ Scaling of Wireless Foundation Models\\ via Representation Diversity and\\ Multi-Branch Architectures}



\author{\IEEEauthorblockN{Ahmed Mohamed, Ahmed Aboulfotouh, and Hatem Abou-Zeid}
\IEEEauthorblockA{{Department of Electrical and Software Engineering}, 
{University of Calgary}, Canada}}

\maketitle

\begin{abstract}

Wireless foundation models learn representations from unlabeled radio signals for reuse across downstream tasks. Scaling model capacity is a common strategy for learning richer representations and improving downstream performance. However, its gains are less consistent in wireless self-supervised learning when pretraining data are limited. We investigate objective diversity as an alternative scaling axis: different self-supervised objectives emphasize different signal properties, and combining their representations can preserve more information to enable diverse tasks. We develop a fusion framework that combines frozen representations from independent encoders trained through reconstructive, predictive, and contrastive learning. This provides a reference for the benefits of diversity, but requires the maintenance of multiple encoders. To retain these benefits within the parameter budget of a standard single-objective encoder, we introduce a jointly trained multi-branch architecture with a shared trunk and objective-specific branches. We pretrain on a heterogeneous corpus of spectrogram and channel state information data and evaluate six downstream tasks that span communication, sensing, and positioning. At an equal output embedding dimension, fusion outperforms the evaluated single-objective encoders on all six tasks while using roughly one-third of their parameters. The multi-branch architecture retains much of the fusion benefit within a single-encoder parameter budget. Representation analyses indicate complementary contributions across objectives, with much of the added benefit retained in components orthogonal to the reconstruction representation subspace. These findings support objective diversity as an effective strategy for scaling wireless foundation models.
\end{abstract}


\section{Introduction}

Foundation models pretrained with self-supervised learning have become the standard approach for building general-purpose representations in vision, language, and speech~\cite{mae, bert, hubert}. The wireless domain has begun to adopt this paradigm, with recent work training transformer encoders on unlabeled radio signals and transferring them to a diverse set of tasks such as signal classification, localization, and channel prediction~\cite{wavesfm, wifo, lwm, latentwave, fastwfms}. These models promise a single backbone that serves many downstream applications without task-specific pretraining. Achieving this requires representations that retain the different signal properties needed by each downstream task, even when those tasks are not known during pretraining.

Progress in other domains has largely come from scaling: larger models trained on larger datasets produce better representations~\cite{scalingnlp, scalingvit, gpt3, scalingspeech, whisper}. Applying this recipe to wireless is challenging: public corpora remain small and fragmented relative to web-scale text and image datasets. Moreover, differences in signal modalities, antenna configurations, and sampling rates further complicate their aggregation. Increasing model capacity alone also leaves the learning objective unchanged: a larger encoder may become better at solving its pretraining task without retaining a broader range of information useful for downstream tasks. Prior work on wireless self-supervised learning has shown that increasing model size does not necessarily improve downstream performance~\cite{radiofm}, a trend that we comprehensively examine and confirm in this work. This raises a central question: \emph{how can additional capacity be used more effectively when pretraining data are limited?}

Instead of adding capacity to one objective, we investigate \emph{objective diversity} as an alternative axis of scaling. Different self-supervised techniques impose different inductive biases through what they require the encoder to learn: masked reconstruction encourages learning fine-grained local structure, latent prediction favors learning higher-level spatial dependencies, and contrastive learning captures view-invariant global features~\cite{mae, jepa, simclr}. These objectives emphasize different properties of the same signals, and information that is unnecessary for one objective may still be useful for a downstream task. Combining their representations could therefore preserve a broader range of useful features than any single objective alone. 

The value of retaining these complementary features is particularly clear in wireless, where downstream tasks depend on different signal properties. For example, representations that preserve temporal and spectral characteristics of the transmitted waveform are useful for modulation classification, whereas those that retain phase relationships across antennas support angle-of-arrival estimation~\cite{iqfm}. Information that a representation suppresses as irrelevant to one task may therefore be essential to another. We hypothesize that combining representations learned under complementary pretraining objectives yields stronger downstream performance than allocating the same capacity to a single-objective encoder.

We develop a representation fusion framework that brings together representations learned independently under multiple self-supervised objectives. This provides an empirical reference for the value of objective diversity, but ties its benefits to the cost of maintaining separate encoders. To reduce this cost, we introduce a multi-branch architecture that combines shared feature learning with objective-specific specialization. A shared trunk learns features across objectives, while separate branches allow their representations to specialize. This design realizes objective diversity within a single jointly trained model, matching the size budget of a standard single-objective encoder.

We instantiate the framework with three objectives spanning reconstructive, predictive, and contrastive learning: MAE, JEPA, and SimCLR, respectively. Evaluation across six downstream tasks spanning spectrogram and CSI modalities supports objective diversity as an effective alternative to single-objective model scaling in the evaluated setting.


\subsection{Contributions}

This work introduces a framework for learning diverse wireless representations through complementary self-supervised objectives and realizing their benefits within the parameter budget of a single encoder. Our contributions are summarized as follows:
\begin{itemize}
\item We propose a fusion framework that combines representations from three complementary self-supervised objectives. Pairwise comparisons show that the strongest objective combination varies across downstream tasks, highlighting the value of retaining features learned under different objectives.
\item We provide evidence for representation diversity as an alternative scaling axis through systematic comparisons with single-objective encoders. Under limited pretraining data, three-objective fusion outperforms single-objective encoders with a matched output embedding dimension on all six tasks, while using roughly one-third of their parameters.
\item We propose a multi-branch architecture that realizes objective diversity through a shared trunk of early transformer blocks and objective-specific branches in the deeper layers. This design recovers most of the fusion gains within the size budget of a single standard encoder, with performance remaining stable across the evaluated shared-depth configurations.
\item We conduct an extensive performance evaluation to analyze how the representations learned under different objectives contribute to the gains from fusion. Centered Kernel Alignment reveals structural differences among the encoders, and probe-weight and leave-one-out diagnostics indicate complementary contributions from all three objectives. Subspace decomposition further shows that much of the benefit of adding other objectives to MAE is retained in components orthogonal to its representation subspace.
\end{itemize}

The remainder of this paper is organized as follows. Section~\ref{sec:related} situates our work within wireless foundation models and self-supervised learning. Section~\ref{sec:methodology} presents the fusion framework and the multi-branch architecture. Section~\ref{sec:results} reports downstream results across the six tasks and provides the representation analysis. Section~\ref{sec:conclusion} concludes our work.


\subsection{Notation}
\label{sec:notation}

Scalars are denoted by plain letters, e.g., $x$; vectors by bold lowercase letters, e.g., $\mathbf{x}$; matrices by bold uppercase letters, e.g., $\mathbf{X}$; and sets by calligraphic letters, e.g., $\mathcal{X}$. The operator $\operatorname{concat}(\cdot)$ denotes feature concatenation, and $|\theta|$ denotes the number of scalar parameters in $\theta$. Table~\ref{tab:notation} summarizes the notation used throughout the paper.

\begin{table}[t]
    \caption{Notation Summary}
    \label{tab:notation}
    \centering
    \footnotesize
    \renewcommand{\arraystretch}{1.15}
    \setlength{\extrarowheight}{2pt}
    \begin{tabular}{@{}p{0.28\columnwidth}p{0.66\columnwidth}@{}}
        \hline
        \textbf{Symbol} & \textbf{Definition} \\
        \hline
        $\mathbf{x},\,\mathcal{X}$
        & Input signal and input space \\
        $\mathcal{D}_u,\,N_u$
        & Unlabeled pretraining dataset and sample count \\
        $C,\,H,\,W$
        & Input channel count, height, and width \\
        $p_h,\,p_w$
        & Patch height and width \\
        $N$
        & Number of input patch tokens \\
        $M,\,m$
        & Number of pretraining objectives and objective index \\
        $\mathcal{L}_m,\,\lambda_m$
        & Pretraining loss for objective $m$ and its weight \\
        $E_{\theta_m}$
        & Independent encoder trained under objective $m$ \\
        $d,\,D$
        & Individual and concatenated representation dimensions,
          with $D=Md$ \\
        $\mathbf{z}_m$
        & Pooled representation for objective $m$ \\
        $\mathbf{z}_{\mathrm{fuse}},\,\mathbf{z}_{\mathrm{mb}}$
        & Concatenated-fusion and multi-branch representations \\
        $L_0$
        & Number of blocks in the baseline encoder \\
        $L_s,\,L_b$
        & Number of shared blocks and blocks per branch \\
        $E_{\text{shared}}$
        & Shared-trunk encoder \\
        $E_{\theta_m}^{\mathrm{branch}}$
        & Objective-specific branch encoder \\
        $P_{\mathrm{base}},\,P_{\mathrm{fuse}},\,P_{\mathrm{mb}}$
        & Baseline, fusion, and multi-branch parameter counts \\
        $K,\,k$
        & Number of downstream tasks and task index \\
        $\mathcal{D}_k,\,g_{\phi_k}$
        & Labeled dataset and prediction head for task $k$ \\
        $B$
        & Pretraining batch size \\
        $\tau_{\mathrm{EMA}},\,\tau_{\mathrm{con}}$
        & Exponential moving average decay and contrastive temperature \\
        \hline
    \end{tabular}
\end{table}

\section{Related Work}
\label{sec:related}
Self-supervised pretraining underpins nearly all wireless foundation models (WFMs), and existing work almost universally commits to a single objective family. The masked-reconstruction line is the most developed. WavesFM~\cite{wavesfm} pretrains a Vision Transformer with masked wireless modeling over spectrograms, CSI, and OFDM resource grids, transferring to 5G NR positioning, channel estimation, human activity sensing, and signal classification through shared parameters and lightweight task heads. WiFo~\cite{wifo} applies a masked autoencoder to space-time-frequency CSI, frames channel prediction as masked reconstruction, and evaluates zero-shot generalization across heterogeneous antenna, subcarrier, and time configurations. The multimodal WFM of~\cite{mmwfm} extends masked wireless modeling to process raw IQ and image-like modalities within one backbone. WiMAE, the reconstruction-only foundation underlying ContraWiMAE~\cite{contrawimae}, similarly builds a masked autoencoder for multi-antenna channel matrices.

\begin{figure*}[t]
    \centering
    \includegraphics[width=\linewidth]{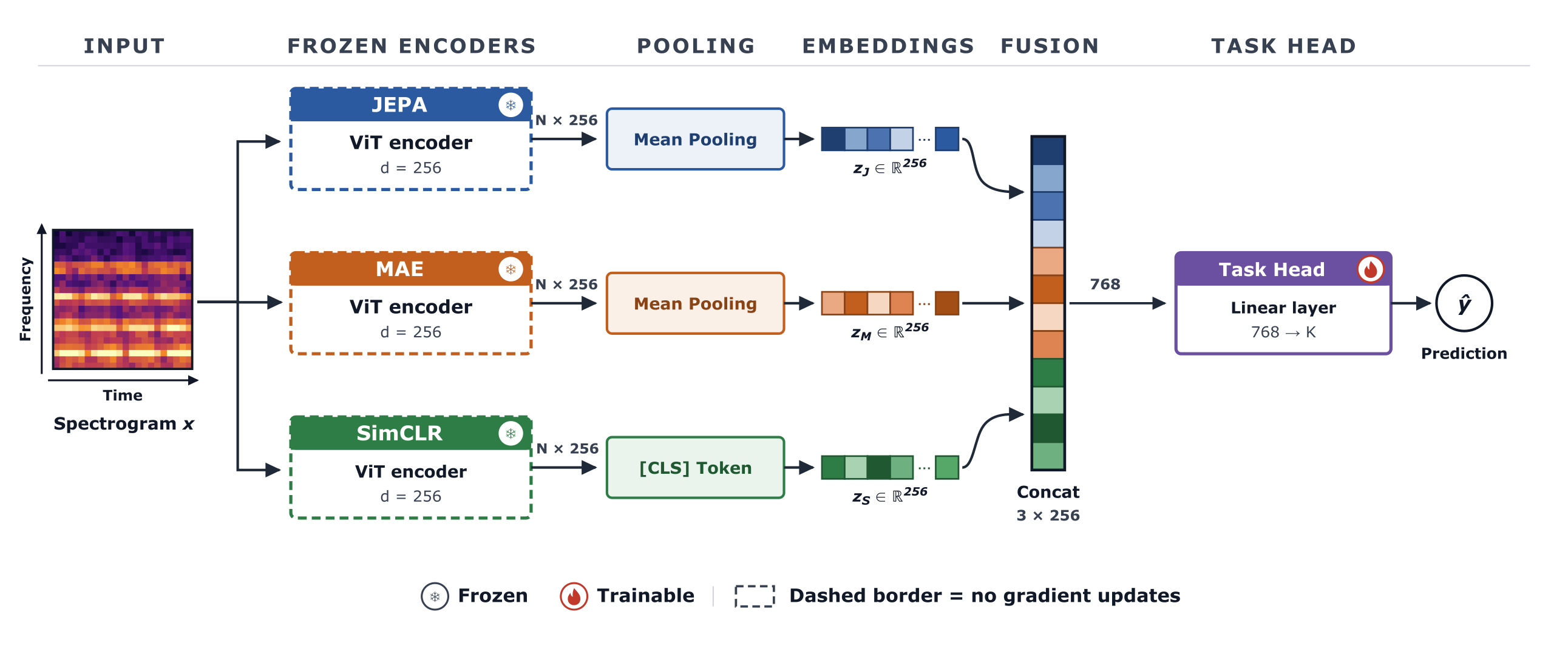}
    \caption{Representation fusion framework. Three encoders, each pretrained with a distinct self-supervised objective, are frozen and their representations are pooled (mean pooling for MAE and JEPA; \texttt{[CLS]} token for SimCLR) and concatenated to form the fused embedding $\mathbf{z}^{\text{fused}}$.}
    \label{fig:fusion}
\end{figure*}

A second, fast-growing line replaces raw reconstruction with prediction in latent space. LatentWave~\cite{latentwave} adapts JEPA to wireless spectrograms and CSI, predicting masked-region representations rather than pixels, and shows that masking geometry induces a task-dependent inductive bias. WirelessJEPA~\cite{wirelessjepa} applies convolutional encoders within a latent-prediction framework to multi-antenna IQ data, outperforming contrastive approaches on several downstream tasks, though both its pretraining and evaluation remain confined to the IQ domain. Homomorphic world models~\cite{naomis} push latent prediction further, learning action-conditioned CSI dynamics with structured transition operators. A parallel contrastive line learns invariances directly: IQFM~\cite{iqfm} applies contrastive SSL to raw IQ streams, and CSI2Vec~\cite{csi2vec} learns CSI embeddings for positioning and channel charting and validates them across antenna counts and subcarrier allocations without CSI reconstruction or ground-truth positions.

Combining objectives to obtain richer representations is established outside wireless. CAV-MAE~\cite{cavmae} shows masked reconstruction and contrastive learning are complementary in audio-visual SSL, capturing structure that neither objective recovers alone. Within wireless, this direction remains largely unexplored. ContraWiMAE~\cite{contrawimae} augments a masked autoencoder with a masked contrastive objective that exploits noise and fading as natural augmentation, reporting gains over reconstruction alone on beam selection, LoS detection, and channel estimation. Realizing objective diversity efficiently requires sharing computation: hard parameter sharing, where a common trunk feeds task-specific heads, is a long-standing multi-task learning technique~\cite{caruana}, and later work learns where to share versus branch rather than fixing the split by hand~\cite{branch_mtl}. Shared-encoder architectures such as CAV-MAE apply the same principle to self-supervised objectives, though previous branching designs target supervised multi-task learning or multiple augmented views of a single objective.

Our work departs from the single-objective paradigm and treats objective diversity, rather than model size, as the primary scaling axis. We differ from prior work in three respects. First, we fuse three objectives spanning all three SSL families, reconstructive (MAE), predictive (JEPA), and contrastive (SimCLR), whereas prior wireless work combines at most two. Second, we frame representation diversity explicitly as a scaling axis and compare it head-to-head against width scaling under a matched parameter budget. Third, our multi-branch design factors a single encoder into a shared trunk and objective-specific branches, recovering the bulk of the fusion gains at zero parameter overhead relative to a single-objective encoder, and we give a mechanistic account of why fusion helps in terms of the structural uniqueness of the encoders and the orthogonality of their complementary information.

\begin{figure*}[t]
    \centering
    \includegraphics[width=\linewidth]{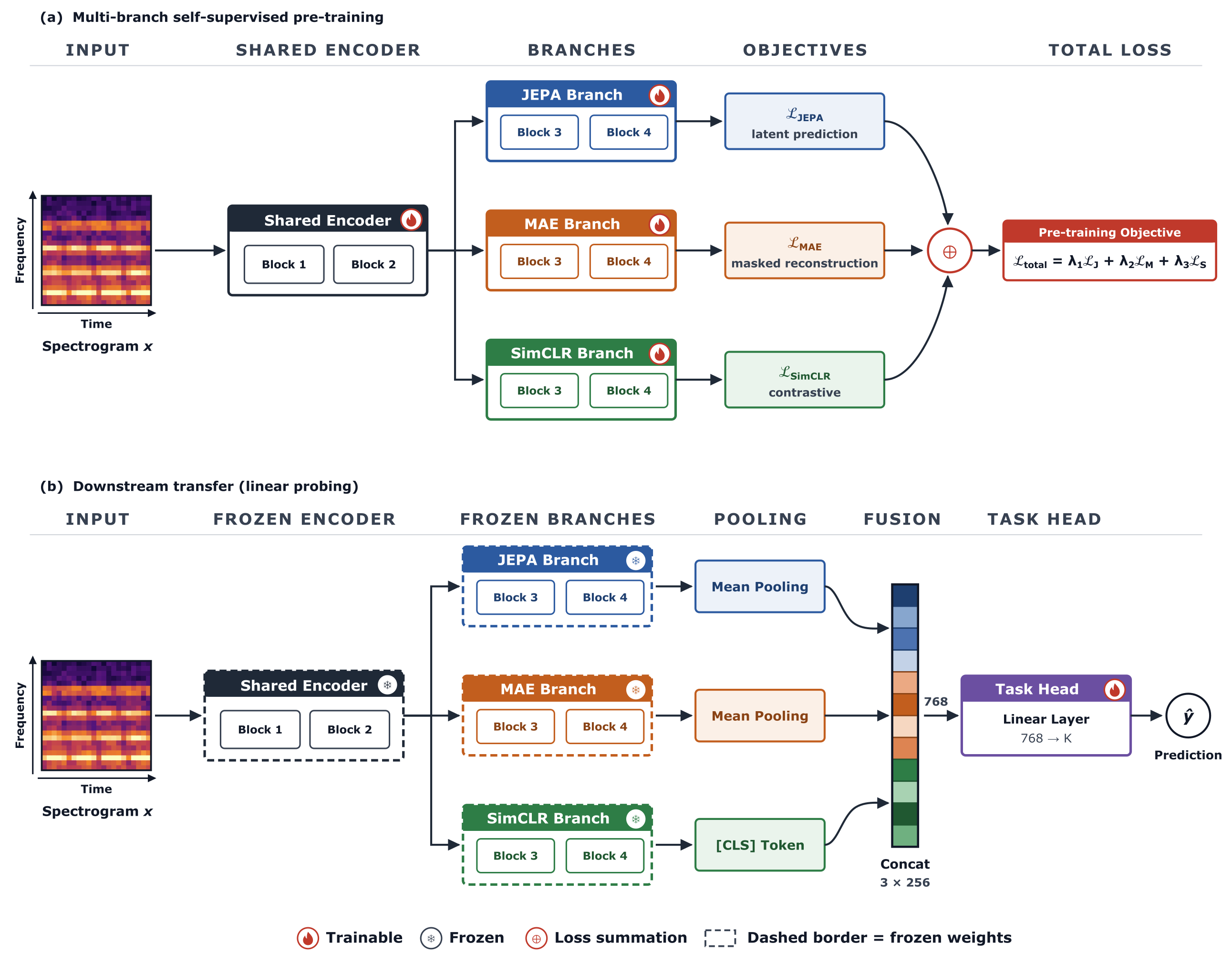}
    \caption{Multi-branch architecture. A shared trunk feeds three branches, one per objective (MAE, JEPA, SimCLR), which are pretrained together using a weighted sum of their losses. For downstream tasks, the pooled branch outputs are concatenated. The model uses the same number of parameters as a single 8-block encoder.}
    \label{fig:multibranch}
\end{figure*}

\section{Methodology}
\label{sec:methodology}

This section develops our approach to scaling wireless foundation models through representation diversity. We begin by formalizing the problem and our hypothesis (Section~\ref{sec:problem}). We then present two instantiations of this idea: a representation fusion framework that trains three independent encoders and combines their outputs (Section~\ref{sec:fusion}) - followed by a \emph{multi-branch} architecture that achieves comparable diversity within the size budget of a single encoder (Section~\ref{sec:multibranch}). Finally, we describe the heterogeneous pretraining corpus (Section~\ref{sec:pretraining}) and the suite of six downstream tasks used to evaluate the learned representations (Section~\ref{sec:downstream}).

\subsection{Problem Formulation}
\label{sec:problem}

Consider a wireless foundation model composed of an encoder $E_{\boldsymbol{\theta}}: \mathcal{X} \rightarrow \mathbb{R}^d$ that maps an input signal $\mathbf{x} \in \mathcal{X}$ to a latent representation $\mathbf{z} = E_{\boldsymbol{\theta}}(\mathbf{x})$ of dimension $d$, where $\boldsymbol{\theta}$ denotes the model parameters. The encoder is pretrained on an unlabeled dataset $\mathcal{D}_u = \{\mathbf{x}_i\}_{i=1}^{N}$ by minimizing a self-supervised objective $\mathcal{L}$, and subsequently evaluated on $K$ downstream tasks via linear probing, where a task-specific head $g_k: \mathbb{R}^d \rightarrow \mathcal{Y}_k$ is trained on a labeled subset $\mathcal{D}_k = \{(\mathbf{x}_j, y_j)\}_{j=1}^{N_k}$ while $\boldsymbol{\theta}$ remains frozen.

In the conventional scaling paradigm, one increases the representation capacity by growing $d$ (and proportionally the number of parameters $|\boldsymbol{\theta}|$) under a fixed objective $\mathcal{L}$. Let $\mathcal{P}_{k}(d, \mathcal{L})$ denote the downstream performance on task $k$ as a function of embedding dimension and pretraining objective. Empirically, we observe diminishing returns:
\begin{equation}
    \frac{\partial \, \mathcal{P}_k(d, \mathcal{L})}{\partial \, d} \rightarrow 0 \quad \text{as } d \text{ grows},
\end{equation}
which we attribute to the limited size and diversity of available wireless pretraining corpora that cannot sustain larger model capacity under a single objective.

We hypothesize that greater downstream gains can be achieved by diversifying the pretraining objective rather than scaling the model. Specifically, given $M$ distinct self-supervised objectives $\{\mathcal{L}_m\}_{m=1}^{M}$, each inducing an encoder $E_{\boldsymbol{\theta}_m}: \mathcal{X} \rightarrow \mathbb{R}^{d_m}$, we define the fused representation as:
\begin{equation}
    \label{eq:z_fused}
    \mathbf{z}^{\text{fused}} = \bigoplus_{m=1}^{M} E_{\boldsymbol{\theta}_m}(\mathbf{x}) \in \mathbb{R}^{D}, \quad D = \sum_{m=1}^{M} d_m,
\end{equation}
where $\bigoplus$ denotes concatenation. Our central claim is that, for a fixed dataset size, representation diversity provides a more effective scaling axis:
\begin{equation}
    \mathcal{P}_k\!\left(D,\, \{\mathcal{L}_m\}_{m=1}^{M}\right) \;>\; \mathcal{P}_k\!\left(D,\, \mathcal{L}\right), \quad \forall\, k,
\end{equation}
i.e., a $D$-dimensional representation composed of $M$ diverse sub-representations outperforms a $D$-dimensional representation trained under any single objective $\mathcal{L}$.

The remaining challenge is to retain the benefits of fusion within the parameter budget of one constituent encoder, avoiding the cost of maintaining \(M\) independent encoders.

\subsection{Fusion Framework}
\label{sec:fusion}

We instantiate the diversity framework of Section~\ref{sec:problem} with $M=3$ complementary self-supervised objectives as illustrated in Fig.~\ref{fig:fusion}: a Masked Autoencoder (MAE)~\cite{mae, wavesfm}, which reconstructs masked input patches; a Joint Embedding Predictive Architecture (JEPA)~\cite{jepa, latentwave}, which predicts masked representations in latent space; and a contrastive method (SimCLR)~\cite{simclr}, which enforces invariance across augmented temporal views. Each objective induces an encoder $E_{\boldsymbol{\theta}_m}$, $m \in \{\text{MAE}, \text{JEPA}, \text{SimCLR}\}$, that is pretrained independently on the unlabeled corpus $\mathcal{D}_u$ and subsequently frozen. 
Each encoder $E_{\boldsymbol{\theta}_m}$ is a \ac{vit}~\cite{vits} with $L=8$ blocks and hidden dimension $d=256$, producing a fixed-dimensional representation through mean pooling of its final token sequence for MAE and JEPA, or extraction of the learned \texttt{[CLS]} token for SimCLR. The three output vectors are then concatenated according to Eq.~\eqref{eq:z_fused}, yielding a fused representation of dimension $D=3d=768$. This framework combines representations learned under different objectives at the cost of maintaining three independent encoders. In the following, we detail the common input representation and each pretraining objective.

\subsubsection{Input Representation}

Let $\mathbf{x} \in \mathbb{R}^{C \times H \times W}$ denote a wireless input signal, where $C$ is the number of antenna channels, $H$ the number of frequency bins, and $W$ the number of time steps. We follow the same tokenization technique adopted by LatentWave~\cite{latentwave}. Rather than treating the multi-antenna input as a volumetric tensor, we process each antenna channel $\mathbf{x}^{(c)} \in \mathbb{R}^{H \times W}$, $c = 1, \ldots, C$, independently by partitioning it into a grid of $\frac{H}{p_h} \times \frac{W}{p_w}$ non-overlapping patches of spatial extent $p_h \times p_w$. A shared linear projection embeds each single-channel patch into a $D$-dimensional vector, producing a token sequence of length $N = C \cdot \frac{H}{p_h} \cdot \frac{W}{p_w}$:
\begin{equation}
    \mathbf{t} = [t_1, t_2, \ldots, t_N] \in \mathbb{R}^{N \times d}.
\end{equation}
Because the projection operates independently on each antenna channel, the embedding layer is agnostic to the number of antennas, enabling the same architecture to accommodate inputs with varying $C$ without structural modification. In addition, the encoder can model the interactions between different tokens on different channels because they are tokenized as a single sequence. Sinusoidal positional embeddings are added to each token to encode its spatial location within the frequency--time grid.

To further decouple the model from any fixed antenna configuration, we adopt a stochastic channel sampling strategy during pretraining. For each training sample with $C$ available channels, we draw $C' \sim \mathcal{U}\{1, C\}$ uniformly at random and retain an arbitrary subset of $C'$ channels, discarding the remainder. The effective sequence length thus varies as $N' = C' \cdot \frac{H}{p_h} \cdot \frac{W}{p_w}$ across iterations, exposing the model to diverse multi-antenna configurations and preventing it from relying on a specific channel count.

\subsubsection{Masked Autoencoder (MAE)}

MAE learns fine-grained signal representations by reconstructing masked portions of the input in the original signal space~\cite{mae, wavesfm}. Following~\cite{wavesfm}, we apply an aggressive random masking ratio of 80\%: given the token sequence $\mathbf{z}$, a binary mask $\mathbf{m} \in \{0,1\}^N$ is sampled uniformly at random such that $|\mathcal{M}| = \lfloor 0.8N \rfloor$ tokens are masked, where $\mathcal{M} = \{i : m_i = 1\}$. Only the visible tokens $\mathbf{t}_{\text{vis}} = \{t_i : m_i = 0\}$ are passed through the \ac{vit} encoder, and a lightweight \ac{vit} decoder $D_{\phi}$ with 4 blocks and 128 dimensions reconstructs the masked patches from the resulting latent representations. The objective minimizes the mean squared error over masked positions:
\begin{equation}
    \mathcal{L}_{\text{MAE}} = \frac{1}{|\mathcal{M}|} \sum_{i \in \mathcal{M}} \left\| \mathbf{p}_i - \hat{\mathbf{p}}_i \right\|_2^2,
\end{equation}
where $\mathbf{p}_i$ is the original patch and $\hat{\mathbf{p}}_i$ is its reconstruction. The high masking ratio forces the encoder to build rich, context-dependent representations from a sparse set of visible tokens, encouraging it to capture signal structures.

\subsubsection{Joint Embedding Predictive Architecture (JEPA)}

In contrast to MAE's input-space reconstruction, JEPA formulates self-supervision as a prediction task in latent space~\cite{jepa, latentwave}. Following~\cite{latentwave}, we sample $Q=6$ contiguous rectangular region masks $\{\mathcal{M}_1, \ldots, \mathcal{M}_Q\}$ from the patch grid rather than applying random token-level masking. Each mask $\mathbf{m}_q$ corresponds to a spatially coherent block of tokens in the frequency--time plane, and the union of all masked regions defines the prediction targets while the complement forms the context:
\begin{equation}
    \mathbf{t}_{\text{ctx}} = \left\{t_i : i \notin \textstyle\bigcup_{q=1}^{Q} \mathcal{M}_q\right\}.
\end{equation}
The context tokens are processed by the \ac{vit} context encoder $E_{\boldsymbol{\theta}}$ to produce latent context representations, from which a lightweight \ac{vit} predictor network $P_{\boldsymbol{\psi}}$ with 4 blocks and 128 dimensions generates predictions for each masked region. The prediction targets are produced by a target encoder $E_{\bar{\boldsymbol{\theta}}}$ that is an \ac{ema} of the context encoder $E_{\boldsymbol{\theta}}$ which processes the full unmasked input. The objective is:
\begin{equation}
    \mathcal{L}_{\text{JEPA}} = \frac{1}{Q} \sum_{q=1}^{Q} \frac{1}{|\mathcal{M}_q|} \sum_{i \in \mathcal{M}_q} \left\| E_{\bar{\boldsymbol{\theta}}}(\mathbf{t_{full}})_i - P_{\boldsymbol{\psi}}\!\left(E_{\boldsymbol{\theta}}(\mathbf{t_{ctx}})\right)_i \right\|_2^2.
\end{equation}
The use of contiguous region masks forces the model to predict coherent spatial blocks from surrounding context, encouraging it to learn higher-level predictive dependencies. By operating in latent space, JEPA avoids committing to low-level reconstruction details and instead captures abstract structural relationships across the signal.

\subsubsection{Simple Contrastive Learning of Representations (SimCLR)}

SimCLR adopts a fundamentally different paradigm by forgoing masking entirely and instead learning representations that are invariant across different augmented views of the same signal~\cite{simclr}. For each input $\mathbf{x}$, we generate two augmented views $\mathbf{x}_1, \mathbf{x}_2$ by extracting two random temporal crops (i.e., contiguous vertical slices along the time axis $W$) ranging from 30\% to 100\% from the full signal. Both views share the same frequency extent $H$ and antenna channels but cover different (possibly overlapping) time intervals, encouraging the encoder to learn features that are stable across temporal shifts.

Each view is independently processed by the encoder, and the resulting representations are passed through a nonlinear 2-layer MLP projection head $G_{\boldsymbol{\omega}}: \mathbb{R}^d \rightarrow \mathbb{R}^{d_p}$. For a mini-batch of $B$ samples yielding $2B$ projected embeddings $\{\mathbf{g}_1, \mathbf{g}_1', \ldots, \mathbf{g}_B, \mathbf{g}_B'\}$, the NT-Xent loss treats the two views of the same sample as a positive pair and all other $2(B-1)$ embeddings as negatives:
\begin{equation}
    \mathcal{L}_{\text{SimCLR}} = -\frac{1}{2B}\sum_{i=1}^{B} \log \frac{\mathrm{exp}(\mathrm{sim}(\mathbf{g}_i, \mathbf{g}_i')/\tau_c)}{\displaystyle\sum_{\substack{j=1,  j \neq i}}^{2B} \mathrm{exp}(\mathrm{sim}(\mathbf{g}_i, \mathbf{g}_j)/\tau_c)}
\end{equation}
where $\mathrm{sim}(\cdot,\cdot)$ denotes cosine similarity and $\tau_c$ is a temperature parameter. By contrasting temporal crops, SimCLR learns globally discriminative features that capture time-invariant features.

\begin{algorithm}[htbp]
\caption{Multi-Branch Pretraining and Linear Probing Evaluation}
\label{alg:main}
\begin{algorithmic}[1]
\REQUIRE Dataset $\mathcal{D}$, shared blocks $E_{\text{shared}}$, branch blocks $\{E_{\theta_{MAE}}^{\mathrm{branch}}, E_{\theta_{JEPA}}^{\mathrm{branch}}, E_{\theta_{SimCLR}}^{\mathrm{branch}}\}$, target encoder $E_{\text{target}}$, decoder $D_{\text{MAE}}$, predictor $P_{\text{JEPA}}$, projector $G_{\text{SimCLR}}$, loss weights $(\lambda_1, \lambda_2, \lambda_3)$, \ac{ema} range $[\tau_{min}, \tau_{max}]$, epochs $N$, total steps $T$
\REQUIRE Downstream dataset $\mathcal{D}_{\text{down}} = \{(\mathbf{x}_i, y_i)\}$, number of seeds $S$, train ratio $r$

\vspace{0.5em}
\STATE \textbf{--- Phase 1: Pretraining ---}
\vspace{0.3em}
\STATE $t \leftarrow 0$
\FOR{epoch $= 1$ to $N$}
\FOR{each batch $\mathbf{x} \in \mathcal{D}$}
    \vspace{0.5em}
    \STATE $\triangleright$ \textbf{Forward passes}
    \STATE $\mathcal{L}_{\text{MAE}} \leftarrow MAEForward(\mathbf{x})$ \COMMENT{Alg.~\ref{alg:mae}}
    \STATE $\mathcal{L}_{\text{JEPA}} \leftarrow JEPAForward(\mathbf{x})$ \COMMENT{Alg.~\ref{alg:jepa}}
    \STATE $\mathcal{L}_{\text{SimCLR}} \leftarrow SimCLRForward(\mathbf{x})$ \COMMENT{Alg.~\ref{alg:simclr}}

    \vspace{0.5em}
    \STATE $\triangleright$ \textbf{Loss computation and backpropagation}
    \STATE $\mathcal{L} \leftarrow \sum_{k} \lambda_k \cdot \mathcal{L}_k$
    \STATE Backpropagate $\mathcal{L}$; update all parameters

    \vspace{0.5em}
    \STATE $\triangleright$ \textbf{EMA and schedule updates}
    \STATE $\tau_t \leftarrow \tau_{min} + (\tau_{max} - \tau_{min}) \cdot \frac{t}{T}$ \COMMENT{Linear EMA schedule}
    \STATE $E_{\text{target}} \leftarrow \tau_t \cdot E_{\text{target}} + (1 - \tau_t) \cdot [E_{\text{shared}}, E_{\theta_{JEPA}}^{\mathrm{branch}}]$
    \STATE Update learning rate and weight decay \COMMENT{Cosine schedules}
    \STATE $t \leftarrow t + 1$
\ENDFOR
\ENDFOR

\vspace{0.8em}
\STATE \textbf{--- Phase 2: Linear Probing ---}
\vspace{0.5em}
\STATE Freeze $E_{\text{shared}}, E_{\theta_{MAE}}^{\mathrm{branch}}, E_{\theta_{JEPA}}^{\mathrm{branch}}, E_{\theta_{SimCLR}}^{\mathrm{branch}}$

\vspace{0.5em}
\STATE $\triangleright$ \textbf{Feature extraction}
\FOR{each $\mathbf{x}_i \in \mathcal{D}_{\text{down}}$}
    \STATE $\mathbf{h}_{\text{shared}} \leftarrow E_{\text{shared}}(\mathbf{x}_i)$
    \STATE $\mathbf{z}_{\text{MAE}} \leftarrow E_{\theta_{MAE}}^{\mathrm{branch}}(\mathbf{h}_{\text{shared}})$ 
    \STATE $\mathbf{z}_{\text{JEPA}} \leftarrow E_{\theta_{JEPA}}^{\mathrm{branch}}(\mathbf{h}_{\text{shared}})$
    \STATE $\mathbf{z}_{\text{SimCLR}} \leftarrow E_{\theta_{SimCLR}}^{\mathrm{branch}}(\mathbf{h}_{\text{shared}})$
    \vspace{0.5em}
    \STATE $\triangleright$ \textbf{Concatenation along representation dim}
    \STATE $\mathbf{z}_i \leftarrow [\mathbf{z}_{\text{MAE}} \| \mathbf{z}_{\text{JEPA}} \| \mathbf{z}_{\text{SimCLR}}]$
\ENDFOR

\vspace{0.5em}
\STATE $\triangleright$ \textbf{Evaluation over multiple seeds}
\FOR{seed $s = 1$ to $S$}
    \STATE Split $\{(\mathbf{z}_i, y_i)\}$ into $\mathcal{T}_{\text{train}}^{(s)}, \mathcal{T}_{\text{test}}^{(s)}$ with ratio $r : (1-r)$ using seed $s$
    \STATE Fit logistic regression $f^{(s)}$ on $\mathcal{T}_{\text{train}}^{(s)}$
    \STATE $m_s \leftarrow \text{Evaluate}(f^{(s)}, \mathcal{T}_{\text{test}}^{(s)})$ \COMMENT{Evaluation metric on test set}
\ENDFOR

\vspace{0.5em}
\RETURN $\mu = \frac{1}{S}\sum_{s=1}^{S} m_s$, \quad $\sigma = \sqrt{\frac{1}{S-1}\sum_{s=1}^{S}(m_s - \mu)^2}$
\end{algorithmic}
\end{algorithm}

\begin{algorithm}[ht]
\caption{MAEForward}
\label{alg:mae}
\begin{algorithmic}[1]
\REQUIRE Batch $\mathbf{x}$, shared blocks $E_{\text{shared}}$, branch blocks $E_{\theta_{MAE}}^{\mathrm{branch}}$, decoder $D_{\text{MAE}}$

\vspace{0.5em}
\STATE $\triangleright$ \textbf{Randomly mask 80\% of patches}
\STATE $\mathbf{m} \leftarrow \textsc{RandomMask}(\mathbf{x}, \text{ratio}=0.80)$
\STATE $\mathbf{x}_{\text{visible}} \leftarrow \mathbf{x}[\neg \mathbf{m}]$

\vspace{0.5em}
\STATE $\triangleright$ \textbf{Encode visible patches through shared then branch blocks}
\STATE $\mathbf{h} \leftarrow E_{\text{shared}}(\mathbf{x}_{\text{visible}})$
\STATE $\mathbf{z} \leftarrow E_{\theta_{MAE}}^{\mathrm{branch}}(\mathbf{h}, \mathbf{m})$

\vspace{0.5em}
\STATE $\triangleright$ \textbf{Decode and compute reconstruction loss on masked patches}
\STATE $\hat{\mathbf{x}} \leftarrow D_{\text{MAE}}(\mathbf{z})$
\RETURN $\text{MSE}(\hat{\mathbf{x}}[\mathbf{m}],\; \mathbf{x}[\mathbf{m}])$
\end{algorithmic}
\end{algorithm}

\begin{algorithm}[ht]
\caption{JEPAForward}
\label{alg:jepa}
\begin{algorithmic}[1]
\REQUIRE Batch $\mathbf{x}$, shared blocks $E_{\text{shared}}$, branch blocks $E_{\theta_{JEPA}}^{\mathrm{branch}}$, predictor $P_{\text{JEPA}}$, target encoder $E_{\text{target}}$

\vspace{0.5em}
\STATE $\triangleright$ \textbf{Sample six contiguous masks; remainder is context}
\STATE $\{\mathbf{m}_1, \dots, \mathbf{m}_6\} \leftarrow \textsc{SampleContiguousRegions}(\mathbf{x})$
\STATE $\mathbf{x}_{\text{context}} \leftarrow \mathbf{x}[\neg \bigcup_i \mathbf{m}_i]$

\vspace{0.5em}
\STATE $\triangleright$ \textbf{Encode context through shared then branch blocks}
\STATE $\mathbf{h} \leftarrow E_{\text{shared}}(\mathbf{x}_{\text{context}})$
\STATE $\mathbf{z} \leftarrow E_{\theta_{JEPA}}^{\mathrm{branch}}(\mathbf{h})$

\vspace{0.5em}
\STATE $\triangleright$ \textbf{Predict latent targets; target encoder is stop-gradient}
\STATE $\hat{\mathbf{z}} \leftarrow P_{\text{JEPA}}(\mathbf{z}, \{\mathbf{m}_i\})$
\STATE $\mathbf{z}_{\text{target}} \leftarrow E_{\text{target}}(\mathbf{x})$

\vspace{0.5em}
\RETURN $\frac{1}{6} \sum_{i=1}^{6} \text{MSE}(\hat{\mathbf{z}}[\mathbf{m}_i],\; \mathbf{z}_{\text{target}}[\mathbf{m}_i])$
\end{algorithmic}
\end{algorithm}

\begin{algorithm}[ht]
\caption{SimCLRForward}
\label{alg:simclr}
\begin{algorithmic}[1]
\REQUIRE Batch $\mathbf{x}$, shared blocks $E_{\text{shared}}$, branch blocks $E_{\theta_{SimCLR}}^{\mathrm{branch}}$, projector $G_{\text{SimCLR}}$

\vspace{0.5em}
\STATE $\triangleright$ \textbf{Generate two temporal views}
\STATE $\mathbf{x}_1, \mathbf{x}_2 \leftarrow \textsc{TimeCrop}(\mathbf{x}, n_{\text{views}}=2)$

\vspace{0.5em}
\STATE $\triangleright$ \textbf{Encode both views through shared then branch blocks}
\STATE $\mathbf{h}_1 \leftarrow E_{\text{shared}}(\mathbf{x}_1)$
\STATE $\mathbf{h}_2 \leftarrow E_{\text{shared}}(\mathbf{x}_2)$
\STATE $\mathbf{z}_1 \leftarrow E_{\theta_{SimCLR}}^{\mathrm{branch}}(\mathbf{h}_1)$
\STATE $\mathbf{z}_2 \leftarrow E_{\theta_{SimCLR}}^{\mathrm{branch}}(\mathbf{h}_2)$

\vspace{0.5em}
\STATE $\triangleright$ \textbf{Project to contrastive space and compute the loss}
\STATE $\mathbf{g}_1 \leftarrow G_{\text{SimCLR}}(\mathbf{z}_1)$
\STATE $\mathbf{g}_2 \leftarrow G_{\text{SimCLR}}(\mathbf{z}_2)$
\RETURN $\textsc{NT-Xent}(\mathbf{g}_1, \mathbf{g}_2)$
\end{algorithmic}
\end{algorithm}

\subsection{Multi-Branch Architecture}
\label{sec:multibranch}

While the representation fusion of Section~\ref{sec:problem} achieves representation diversity, it requires maintaining $M$ independent encoders, resulting in an $M$-fold increase in overall model size. We now introduce a multi-branch architecture that achieves comparable diversity gains within the same size budget as a single encoder.

The key architectural insight is that early transformer layers tend to learn general, objective-agnostic features (e.g., local patch interactions and low-level signal patterns), whereas deeper layers specialize to the specific pretraining objective. We exploit this by factoring the encoder into a shared trunk followed by objective-specific branches as illustrated in Fig.~\ref{fig:multibranch}. Concretely, let $L$ denote the total number of transformer blocks in a single-method encoder. We partition the $L$ blocks into $L_s$ shared blocks $E_{\text{shared}}$ and $L_b = L - L_s$ branch-specific blocks per objective, yielding the architecture:
\begin{equation}
    E_{\text{shared}}: \mathbb{R}^{N \times d} \rightarrow \mathbb{R}^{N \times d}, \quad E_m: \mathbb{R}^{N \times d} \rightarrow \mathbb{R}^{N \times d},
\end{equation}
for $m \in \{\text{MAE}, \text{JEPA}, \text{SimCLR}\}$. The total parameter count is $L_s + M \cdot L_b$ transformer blocks. Setting $L_s = 2$ and $L_b = 2$ with $M = 3$ yields $2 + 3 \times 2 = 8$ blocks, which matches the $L = 8$ blocks of each single-method encoder. Thus, the multi-branch model introduces \emph{zero} parameter overhead relative to a single dedicated encoder while accommodating three distinct pretraining objectives.

During pretraining, as shown in Algorithm~\ref{alg:main}, the input is passed to each objective to execute its own forward pass. Each objective forward pass follows a four-stage pipeline which is illustrated in Algorithms~\ref{alg:mae},~\ref{alg:jepa}, and~\ref{alg:simclr}: (i) the input is masked or augmented according to the objective's strategy, (ii) the resulting tokens are processed by the shared blocks $E_{\text{shared}}$, (iii) the intermediate output is passed through the objective's dedicated branch blocks, and (iv) the final output is used to compute its loss. At each training step, all three objectives execute this pipeline in parallel on the same input, and the overall training objective is a weighted sum:
\begin{equation}
    \mathcal{L}_{\text{multi}} = \lambda_1 \mathcal{L}_{\text{MAE}} + \lambda_2 \mathcal{L}_{\text{JEPA}} + \lambda_3 \mathcal{L}_{\text{SimCLR}},
\end{equation}
where gradients from all three losses flow back through the branch-specific blocks and into the shared trunk, encouraging $E_{\text{shared}}$ to learn representations that are simultaneously useful for reconstruction, latent prediction, and contrastive invariance.

A subtle design choice concerns the JEPA target encoder. In the independent pretraining setting, the target encoder is an \ac{ema} copy of the full JEPA encoder. In the multi-branch setting, the target encoder must provide representations consistent with the JEPA branch's view of the input, so we construct it as the \ac{ema} of the composed mapping $[E_{\text{shared}}, E_{\theta_{JEPA}}^{\mathrm{branch}}]$:
\begin{equation}
    \boldsymbol{E}_{\text{target}} \leftarrow \tau_t \cdot \boldsymbol{E}_{\text{target}} + (1 - \tau_t) \cdot [\boldsymbol{E}_{\text{shared}},\, \boldsymbol E_{\theta_{JEPA}}^{\mathrm{branch}}],
\end{equation}
where $\tau_t$ follows a linear schedule from $\tau_{\min}$ to $\tau_{\max}$ over the course of training. This ensures that the prediction targets remain aligned with the evolving shared and JEPA-specific parameters.

At inference time, a given input passes through the shared trunk once and is then forwarded through all three branches in parallel. The branch outputs are pooled as in the fusion approach in Section~\ref{sec:fusion} and concatenated to produce the final representation:
Let $\mathbf{z}_m$ denote the pooled output of branch $m$:
\begin{align}
    \mathbf{z}_{\text{MAE}} &= E_{\theta_{MAE}}^{\mathrm{branch}}(\mathbf{h}), \nonumber \\
    \mathbf{z}_{\text{JEPA}} &= E_{\theta_{JEPA}}^{\mathrm{branch}}(\mathbf{h}), \\
    \mathbf{z}_{\text{SimCLR}} &= E_{\theta_{SimcLR}}^{\mathrm{branch}}(\mathbf{h})\nonumber
\end{align}
where $\mathbf{h} = E_{\text{shared}}(\mathbf{x})$. The final representation is then:
\begin{equation}
    \mathbf{z}^{\text{multi}} = \left[\mathbf{z}_{\text{MAE}} \;\|\; \mathbf{z}_{\text{JEPA}} \;\|\; \mathbf{z}_{\text{SimCLR}}\right] \in \mathbb{R}^{3d}.
\end{equation}
With each branch producing a $d$-dimensional vector, the resulting representation has dimension $D = 3d$, identical to the output of the fusion framework.

By co-training three objectives within a size budget of single encoder, the multi-branch architecture internalizes representation diversity as a structural property of the model rather than an emergent property of ensembling independent systems. The shared trunk serves as a common feature extractor that amortizes computation across objectives, while the branch-specific layers provide sufficient capacity for each objective to develop its specialized representation subspace.

\subsection{Pretraining Data}
\label{sec:pretraining}

The pretraining corpus comprises four unlabeled datasets that collectively span both spectrogram and CSI modalities. The first is a proprietary over-the-air spectrogram dataset containing 3200 samples of WiFi, LTE, Bluetooth, 5G NR, and ISM signals, acquired via \ac{sdr} across multiple sub-6\,GHz center frequencies at sampling rates between 10 and 60\,MHz. The second dataset provides approximately 6000 single-channel spectrogram samples collected for RF fingerprinting~\cite{powder}. The third consists of roughly 15000 five-channel samples of indoor 5G NR CSI~\cite{5g_positioning}, and the fourth contains 840 three-channel WiFi CSI samples~\cite{wifi-csi}. In total, the corpus amounts to approximately 25000 samples with heterogeneous antenna configurations ranging from one to five channels.

All spectrogram inputs are converted to decibel scale via the transformation $20\log_{10}(\cdot)$ prior to any further processing. Each sample, regardless of whether it originates as a spectrogram or CSI measurement, is then resized to a spatial resolution of $224 \times 224$ and normalized by subtracting the per-channel mean and dividing by the per-channel standard deviation, both computed over the respective pretraining dataset.

\subsection{Downstream Tasks}
\label{sec:downstream}

We evaluate the learned representations on six downstream tasks that span communication, sensing, and positioning domains. Three tasks operate on CSI inputs and three on spectrogram inputs, providing a comprehensive assessment across both modalities.

\subsubsection{Spectrogram Tasks}

\textbf{RF Signal Classification (RFS).} This task requires identifying the wireless standard of a received signal from its spectrogram. We use the CommRad~RF dataset~\cite{commrad}, which contains over-the-air captures of multiple radio technologies.

\textbf{GNSS Jamming Detection.} This dataset consists of synthesized raw IQ data generated via MATLAB simulation to represent diverse GNSS interference scenarios, including narrowband, AM, FM, chirp, and DME jamming signals~\cite{jamming}. The IQ recordings are converted to spectrograms prior to evaluation, and the task is to classify the jamming type.

\textbf{Interference Classification (INTC).} This task uses over-the-air LTE IQ recordings captured under both clean conditions and in the presence of direct-sequence spread-spectrum (DSSS) interference~\cite{interference}. The IQ samples are transformed into spectrograms, and the objective is to classify the interference modulation type.

\subsubsection{CSI Tasks}

\textbf{5G NR Positioning.} We use the outdoor scenario of the 5G CSI dataset~\cite{5g_positioning}---distinct from the indoor scenario included in the pretraining corpus---to regress the user equipment location from five-channel CSI measurements.

\textbf{Beam Prediction.} Using the DeepMIMO dataset~\cite{deepmimo}, this task predicts the optimal beam index from CSI, framed as a multi-class classification problem.

\textbf{LoS/NLoS Classification.} Also derived from DeepMIMO~\cite{deepmimo}, this binary classification task determines whether the propagation channel between transmitter and receiver contains a line-of-sight path.

All classification tasks are evaluated using per-class accuracy, defined as the mean of each class recall, to account for potential class imbalance. The positioning task is evaluated using mean position error, computed as the average Euclidean distance between predicted and true 3D coordinates.

\section{Results and Discussion}
\label{sec:results}

\begin{table*}[ht]
\centering
\caption{Model configurations for single-objective scaling, fusion, and multi-branch learning. Hidden dimensions are specified per encoder or branch. Parameter counts exclude auxiliary pretraining components and downstream heads.}
\label{tab:model_comparison}
\resizebox{0.9\textwidth}{!}{%
\begin{tabular}{lcccccc}
\toprule
Configuration & Objectives & Models & Block configuration & Hidden dim. & Output dim. & Params (M) \\
\midrule
Single-method (256-d) & 1 & 1 & 8 per encoder & 256 & 256 & 6.4 \\
Single-method (512-d) & 1 & 1 & 8 per encoder & 512 & 512 & 25.5 \\
Single-method (768-d) & 1 & 1 & 8 per encoder & 768 & 768 & 57.1 \\
\midrule
Pairwise fusion      & 2 & 2 & 8 per encoder & 256 & 512 & 12.8 \\
Three-method fusion  & 3 & 3 & 8 per encoder & 256 & 768 & 19.2 \\
\midrule
Multi-branch         & 3 & 1 & 2 shared + 2 per branch & 256 & 768 & 19.2 \\
\bottomrule
\end{tabular}
}
\end{table*}
\begin{table*}[ht]
\centering
\caption{Test performance across downstream tasks (mean $\pm$ std over 20 seeds). Bold indicates best per task.}
\label{tab:main_results}
\resizebox{\textwidth}{!}{%
\begin{tabular}{lcccccc}
\toprule
Configuration & RFS (\%) $\uparrow$ & Positioning (m) $\downarrow$ & Beam (\%) $\uparrow$ & LoS/NLoS (\%) $\uparrow$ & Jamming (\%) $\uparrow$ & Interf. (\%) $\uparrow$ \\
\midrule
\multicolumn{7}{l}{\textit{Single-method (256-d)}} \\
JEPA only          & $80.6 \pm 1.80$ & $2.55 \pm 0.022$ & $54.1 \pm 1.00$ & $92.3 \pm 0.41$ & $87.8 \pm 0.89$ & $75.1 \pm 0.67$ \\
MAE only           & $85.9 \pm 1.57$ & $2.28 \pm 0.024$ & $57.7 \pm 0.84$ & $93.5 \pm 0.36$ & $95.6 \pm 0.42$ & $76.5 \pm 0.58$ \\
SimCLR only        & $84.5 \pm 2.38$ & $3.00 \pm 0.030$ & $50.3 \pm 0.95$ & $92.3 \pm 0.44$ & $91.6 \pm 0.52$ & $72.7 \pm 0.75$ \\
\midrule
\multicolumn{7}{l}{\textit{Single-method (768-d)}} \\
JEPA only          & $79.9 \pm 2.23$ & $2.08 \pm 0.025$ & $62.1 \pm 0.94$ & $93.5 \pm 0.51$ & $88.3 \pm 0.82$ & $83.6 \pm 0.52$ \\
MAE only           & $84.2 \pm 2.05$ & $1.76 \pm 0.021$ & $62.4 \pm 1.05$ & $93.6 \pm 0.41$ & $96.1 \pm 0.39$ & $84.6 \pm 0.65$ \\
SimCLR only        & $86.2 \pm 2.22$ & $2.77 \pm 0.034$ & $54.6 \pm 0.95$ & $93.3 \pm 0.40$ & $94.7 \pm 0.59$ & $76.1 \pm 0.47$ \\
\midrule
\multicolumn{7}{l}{\textit{Fusion (768-d)}} \\
JEPA + MAE + SimCLR & $\mathbf{88.3 \pm 2.00}$ & $\mathbf{1.68 \pm 0.024}$ & $\mathbf{65.2 \pm 0.99}$ & $\mathbf{94.0 \pm 0.46}$ & $\mathbf{96.7 \pm 0.33}$ & $\mathbf{84.8 \pm 0.60}$ \\
\bottomrule
\end{tabular}
}
\end{table*}

\begin{table*}[ht]
\centering
\caption{Pairwise fusion results at 512-d embedding dimension (mean $\pm$ std over 20 seeds). Bold indicates best per task.}
\label{tab:pairwise_results}
\resizebox{\textwidth}{!}{%
\begin{tabular}{lcccccc}
\toprule
Configuration & RFS (\%) $\uparrow$ & Positioning (m) $\downarrow$ & Beam (\%) $\uparrow$ & LoS/NLoS (\%) $\uparrow$ & Jamming (\%) $\uparrow$ & Interf. (\%) $\uparrow$ \\
\midrule
\multicolumn{7}{l}{\textit{Single-method (512-d)}}\\
JEPA only          & $81.0 \pm 1.66$ & $1.98 \pm 0.018$ & $60.0 \pm 0.87$ & $93.1 \pm 0.48$ & $90.8 \pm 0.46$ & $\mathbf{82.2 \pm 0.69}$ \\
MAE only           & $86.5 \pm 2.04$ & $1.96 \pm 0.018$ & $60.7 \pm 0.98$ & $93.8 \pm 0.40$ & $95.9 \pm 0.45$ & $82.1 \pm 0.64$ \\
SimCLR only        & $82.8 \pm 2.66$ & $2.71 \pm 0.031$ & $50.8 \pm 0.94$ & $92.6 \pm 0.42$ & $90.7 \pm 0.49$ & $77.4 \pm 0.69$ \\
\midrule
\multicolumn{7}{l}{\textit{Fusion (512-d)}} \\
JEPA + MAE         & $86.3 \pm 1.79$ & $\mathbf{1.86 \pm 0.026}$ & $\mathbf{62.9 \pm 0.85}$ & $\mathbf{94.0 \pm 0.38}$ & $95.6 \pm 0.40$ & $81.0 \pm 0.59$ \\
JEPA + SimCLR      & $85.7 \pm 2.09$ & $2.13 \pm 0.025$ & $58.5 \pm 1.14$ & $93.1 \pm 0.42$ & $92.7 \pm 0.68$ & $80.7 \pm 0.66$ \\
MAE + SimCLR       & $\mathbf{88.6 \pm 2.06}$ & $1.90 \pm 0.018$ & $61.9 \pm 0.91$ & $\mathbf{94.0 \pm 0.39}$ & $\mathbf{97.0 \pm 0.32}$ & $81.1 \pm 0.64$ \\
\bottomrule
\end{tabular}
}
\end{table*}

We evaluate the learned representations on six downstream wireless tasks and organize our findings around two themes. The first addresses performance: whether representation diversity outperforms single-objective scaling (Section~\ref{sec:fusion_results}), whether this diversity can be realized without parameter overhead (Section~\ref{sec:multibranch_results}), and which pretraining objectives complement each other the most (Section~\ref{sec:complementarity}). The second provides mechanistic insight into why fusion works: we use linear Centered Kernel Alignment (CKA)~\cite{cka} analysis to confirm that the three encoders produce structurally distinct representations (Section~\ref{sec:cka}), show through probe weight and leave-one-out diagnostics that all three objectives contribute non-redundantly to downstream performance (Section~\ref{sec:objective_importance}), and demonstrate via subspace decomposition that the complementary information each encoder provides is orthogonal to the others (Section~\ref{sec:residual}).

\subsection{Representation Diversity versus Single-Objective Scaling}
\label{sec:fusion_results}
We first examine whether combining representations from multiple pretraining objectives improves downstream performance more than increasing the capacity of a single-objective encoder. Table~\ref{tab:model_comparison} summarizes the evaluated configurations and their parameter counts. The comparison includes individual 256-d encoders, their 768-d fusion, and single-objective encoders with a matching 768-d output dimension (Table~\ref{tab:main_results}). Fusion improves over the strongest 256-d baseline (MAE) on all six tasks, including a 7.5 percentage-point gain in beam prediction accuracy and a 0.60\,m reduction in positioning error.

These improvements cannot be explained by the larger output dimension alone. Increasing single-objective encoder width from 256 to 768 dimensions produces gains on some tasks but degrades performance on others. For example, MAE improves on beam prediction and interference classification, but its RFS accuracy falls from 85.9\% to 84.2\%. At the same 768-d output dimension, fusion outperforms all three single-objective encoders on every task while using roughly one-third of their parameters. This parameter advantage arises because fusion combines three smaller encoders, whereas transformer parameter counts grow approximately quadratically with hidden dimension.

Pairwise fusion provides further evidence of this benefit at a 512-d output dimension (Table~\ref{tab:pairwise_results}). The best pairwise fusion per task matches or exceeds the best single-objective encoder on five of the six tasks while using roughly half the parameters; single-objective JEPA retains the best interference accuracy. The strongest objective pair varies across tasks, indicating that the value of combining representations depends on the downstream application. These results support representation diversity as an effective alternative to single-objective scaling under limited pretraining data.

\begin{table*}[ht]
\centering
\caption{Multi-Branch shared depth ablation for JEPA + MAE + SimCLR (768-d, mean $\pm$ std over 20 seeds). $S$ and $B$ denote the number of shared and per-branch blocks, respectively. Overhead is compared to single-objective encoders with 256 dimensions and 8 blocks. It's computed as $(S + 3 \times B - 8) / 8$.}
\label{tab:multi_branch_ablation}
\resizebox{\textwidth}{!}{%
\begin{tabular}{cccccccc}
\toprule
$S$ / $B$ & Overhead & RFS (\%) $\uparrow$ & Positioning (m) $\downarrow$ & Beam (\%) $\uparrow$ & LoS/NLoS (\%) $\uparrow$ & Jamming (\%) $\uparrow$ & Interf. (\%) $\uparrow$ \\
\midrule
6 / 2 & 50.0\% & $86.9 \pm 1.97$ & $1.63 \pm 0.020$ & $62.6 \pm 0.78$ & $92.6 \pm 0.60$ & $94.2 \pm 0.45$ & $86.7 \pm 0.82$ \\
4 / 2 & 25.0\% & $86.6 \pm 2.26$ & $1.63 \pm 0.018$ & $62.1 \pm 0.80$ & $92.5 \pm 0.49$ & $95.9 \pm 0.44$ & $86.8 \pm 0.76$ \\
2 / 2 &  0.0\% & $87.1 \pm 1.81$ & $1.73 \pm 0.023$ & $62.5 \pm 0.86$ & $92.8 \pm 0.37$ & $95.1 \pm 0.51$ & $86.6 \pm 0.55$ \\
\bottomrule
\end{tabular}
}
\end{table*}

\begin{figure}[ht]
    \centering
    \includegraphics[width=\linewidth]{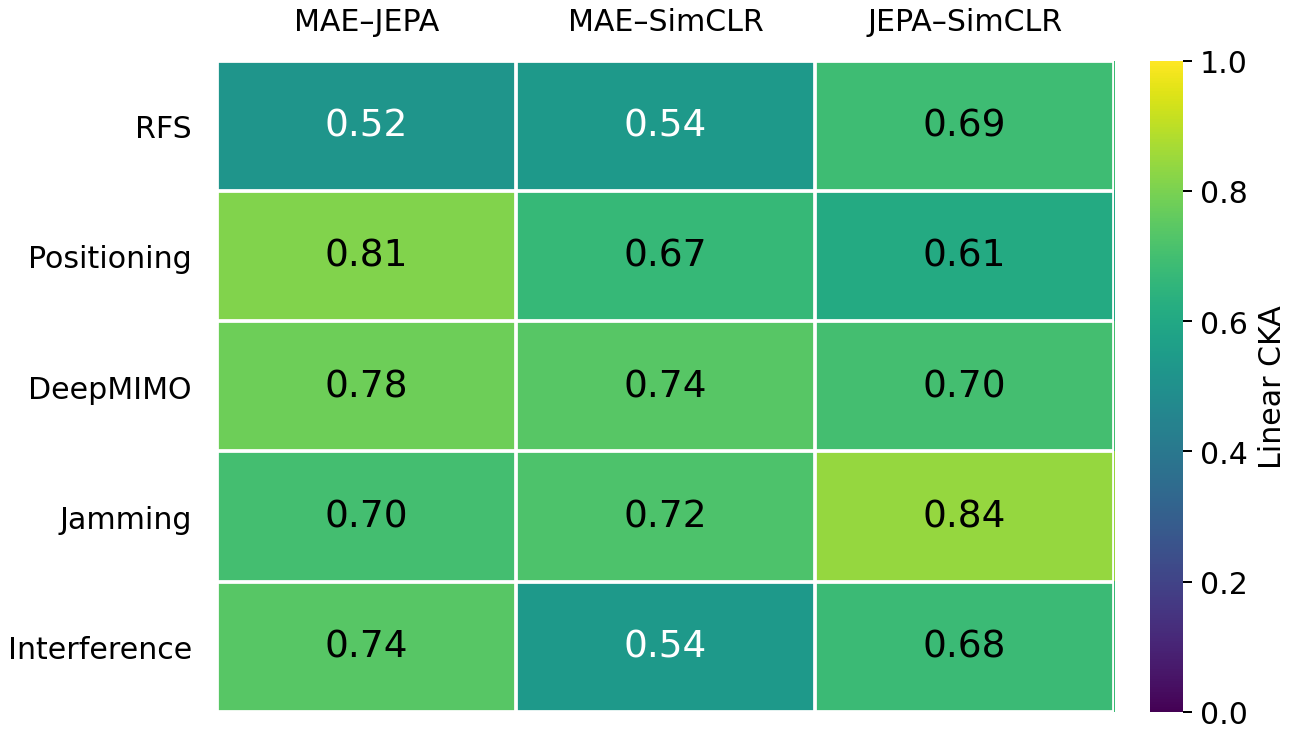}
    \caption{Pairwise linear CKA similarity between representations learned by MAE, JEPA, and SimCLR on each downstream dataset. Lower values indicate greater differences in representation structure. A value of 1 indicates perfect similarity.}
    \label{fig:cka}
\end{figure}

\begin{figure*}[t]
    \centering
    \begin{subfigure}[t]{0.48\linewidth}
        \centering
        \includegraphics[width=\linewidth]{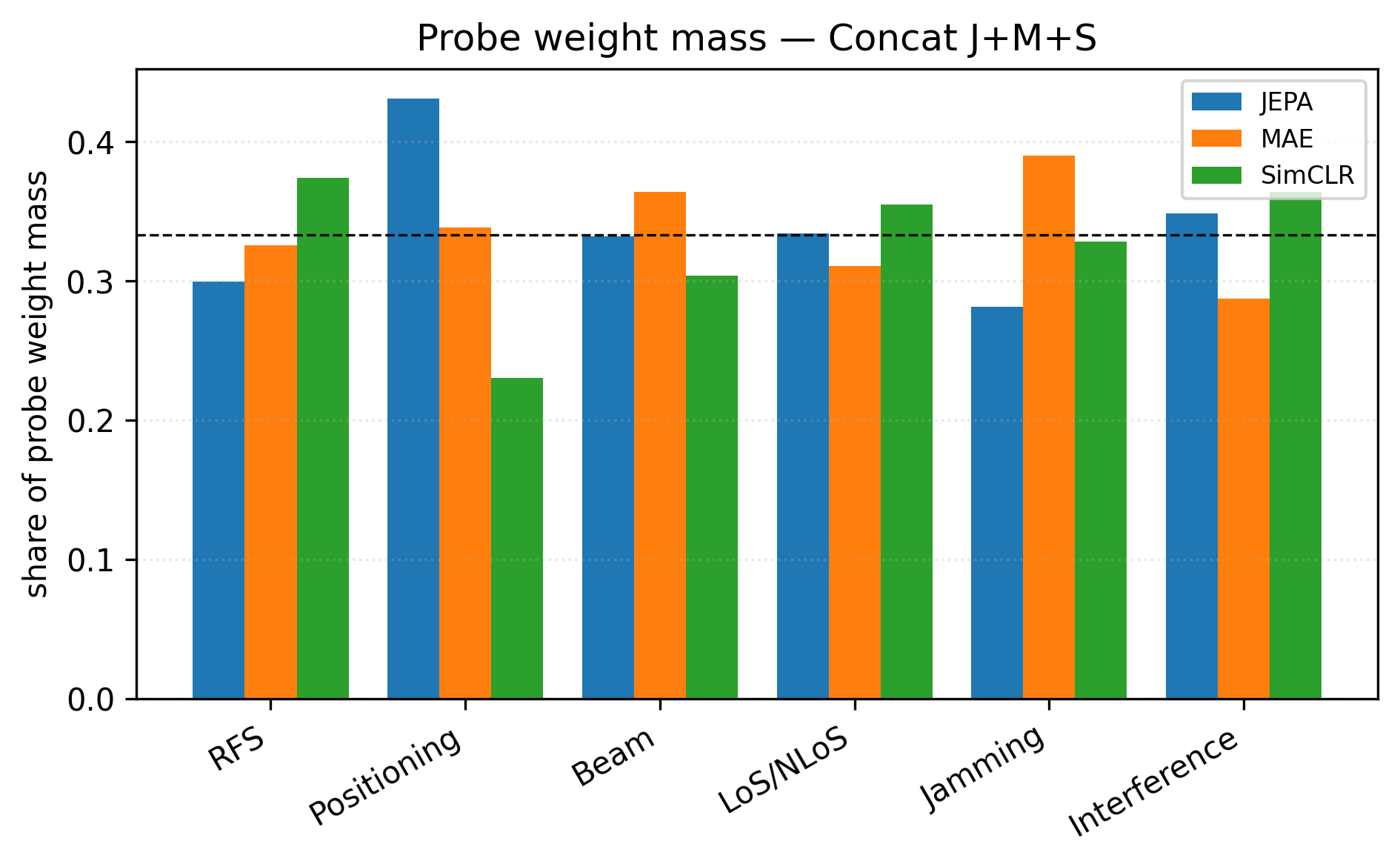}
        \caption{Fraction of total linear probe weight magnitude assigned to each encoder's slice of the concatenated representation. All three objectives receive substantial weight across every task.}
        \label{fig:probe}
    \end{subfigure}
    \hfill
    \begin{subfigure}[t]{0.48\linewidth}
        \centering
        \includegraphics[width=\linewidth]{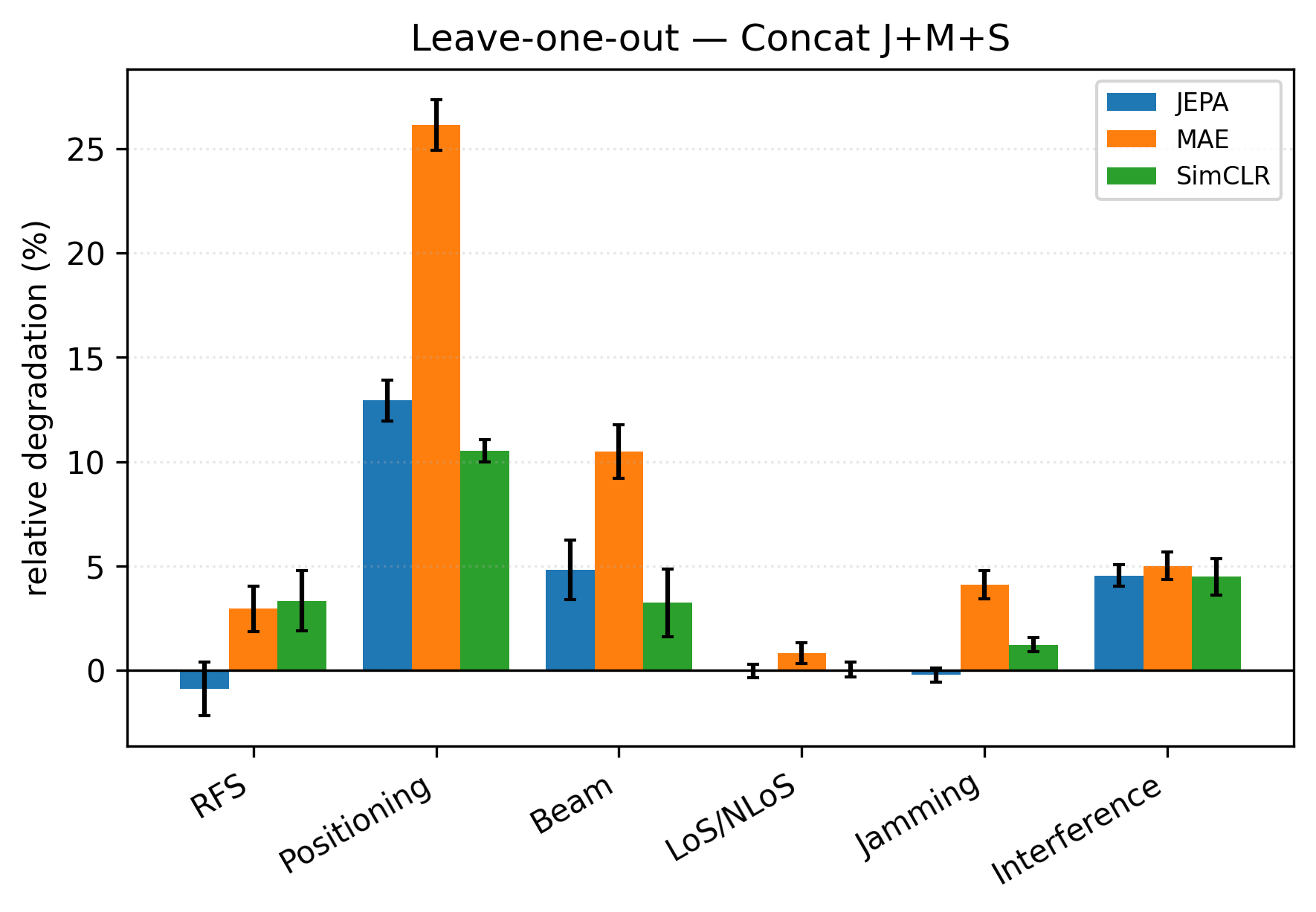}
        \caption{Relative performance drop when one encoder's features are zeroed out. Positive values indicate degradation. Removing any single encoder hurts at least three of the six tasks.}
        \label{fig:lobo}
    \end{subfigure}
    \caption{Objective importance analysis for the fusion. Probe weight mass (left) shows that downstream tasks actively utilize all three encoders, while leave-one-out (right) confirms that the contribution of each encoder is non-redundant.}
    \label{fig:combined}
\end{figure*}

\begin{figure*}[t]
    \centering
    \includegraphics[width=\linewidth]{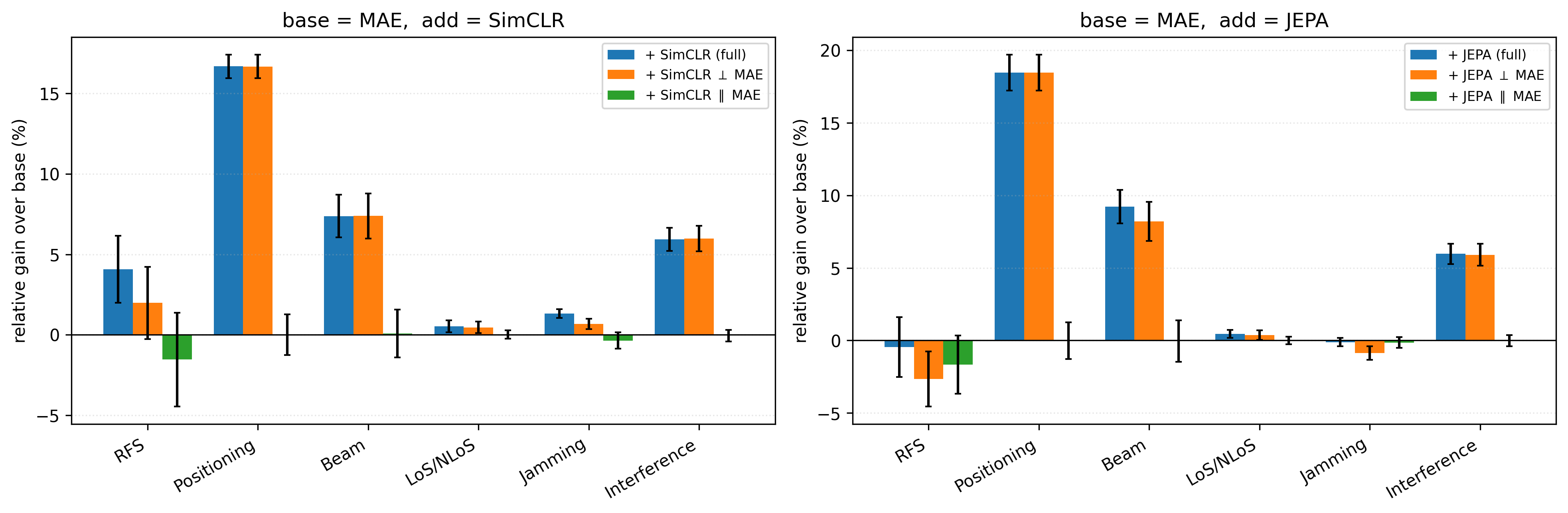}
    \caption{Subspace decomposition of complementary encoder embeddings. We project SimCLR and JEPA embeddings onto MAE's subspace, which splits each one into a parallel component and an orthogonal residual. Each component (full, residual or parallel) is concatenated with MAE and evaluated on all six tasks. The residual alone recovers most or all of the full-embedding gain. The parallel component adds almost nothing and sometimes falls slightly below the MAE-only baseline.}
    \label{fig:residual}
\end{figure*}

\subsection{Multi-Branch Performance across Shared Depths}
\label{sec:multibranch_results}

Fusion triples the parameter count relative to a single constituent encoder. We therefore evaluate the multi-branch alternative introduced in Section~\ref{sec:multibranch}, which combines a shared trunk with objective-specific branches. Table~\ref{tab:multi_branch_ablation} compares configurations $S/B \in \{6/2,\,4/2,\,2/2\}$, where $S$ and $B$ denote the number of shared and per-branch blocks, respectively.

Performance remains similar across these configurations: RFS and beam prediction accuracy each vary by only 0.5 percentage points, while positioning error ranges from 1.63 to 1.73\,m. This stability suggests that two per-branch blocks provide sufficient scope for objective-specific specialization in the evaluated setting. Increasing shared depth to $S=6$ adds 50\% parameter overhead without consistent improvement, suggesting limited sensitivity to shared depth over the tested range.

The zero-overhead configuration ($S=2,\,B=2$) matches the parameter budget of a single 256-d, eight-block encoder while producing a 768-d representation. It achieves higher mean performance than the best single-method 768-d encoder on four of six tasks, despite using roughly one-ninth of its parameters, with the largest accuracy gain on interference classification (+2.0 percentage points). Performance is lower on LoS/NLoS detection and jamming classification, suggesting a possible trade-off when allocating a fixed parameter budget across objectives. Compared with fusion (Table~\ref{tab:main_results}), classification accuracy remains within 2.7 percentage points and positioning error increases by only 0.05\,m, while interference accuracy improves. These results show that a shared trunk with objective-specific branches can retain substantial benefits of representation diversity within a single-encoder parameter budget.

\subsection{Pairwise Complementarity}
\label{sec:complementarity}

Table~\ref{tab:pairwise_results} disaggregates the three-way fusion into its constituent pairs, revealing which objectives are most complementary and for which tasks. MAE + SimCLR emerges as the strongest pair overall, achieving the highest RFS (88.6\%) and Jamming accuracy (97.0\%) among all 512-d configurations, including the single-method baselines. JEPA + MAE, on the other hand, dominates spatial and predictive tasks, delivering the best Positioning error (1.86\,m) and Beam accuracy (62.9\%). JEPA + SimCLR is consistently the weakest pair. One possible explanation is that MAE's reconstructive objective preserves signal details that complement the features learned through prediction or contrastive learning. This interpretation is consistent with MAE's strong standalone performance across the evaluated tasks (Table~\ref{tab:main_results}).

\subsection{Representation Similarity across Objectives}
\label{sec:cka}

To assess whether the three pretraining objectives produce structurally different representations, we compute linear Centered Kernel Alignment (CKA)~\cite{cka} between every encoder pair on each downstream dataset. Figure~\ref{fig:cka} reports the pairwise similarities. Across all five datasets, cross-method CKA values fall substantially below~1, ranging from 0.52 (MAE--JEPA on RFS) to 0.84 (JEPA--SimCLR on Jamming), confirming that the three objectives do not converge to the same embedding geometry. MAE consistently exhibits the lowest similarity to the other two encoders, suggesting it captures the most distinctive feature subspace, which is an observation consistent with its strong standalone performance in Table~\ref{tab:main_results}. JEPA and SimCLR, while more similar to each other on certain tasks (e.g., CKA\,=\,0.84 on Jamming), still maintain moderate divergence on others (e.g., CKA\,=\,0.61 on Positioning), indicating that even the most aligned pair encodes partially different structure depending on the data domain. These results indicate that the three objectives learn different representation structures and do not converge to a common embedding geometry.

\subsection{Contributions of Individual Objectives} 
\label{sec:objective_importance}

Section~\ref{sec:cka} provided evidence that the three encoders learn distinct representation structures. We now examine whether downstream probes exploit these differences and how each objective's contribution varies across tasks. We use two complementary diagnostics applied to the fused representation: probe weight mass and leave-one-out analysis (Figure~\ref{fig:combined}).

\paragraph{Probe weight mass.} For each task, we compute the fraction of the linear probe's total weight magnitude assigned to each encoder's slice of the 768-d concatenated representation. If the probe ignored an encoder, its weight mass would collapse to zero. Figure~\ref{fig:probe} shows substantial weight on all three slices across every task, with no share falling below 23\%. The distribution varies across tasks: JEPA receives the largest share on Positioning (43.1\%), while MAE receives the largest share on Jamming (39.0\%). These patterns suggest that the probes draw on features from all three objectives with task-dependent emphasis. These weight patterns confirm that the downstream probes actively rely on features from all three objectives, not just MAE.

\paragraph{Leave-one-branch-out.} Weight mass shows utilization but cannot distinguish whether the features each encoder provides are replaceable by the remaining two. We therefore zero out one encoder's slice and re-evaluate. A large performance drop indicates non-redundant information. Figure~\ref{fig:lobo} shows that removing MAE causes the largest degradation on every task, including a 26.2\% relative increase in positioning error and a 10.5\% relative drop in beam prediction accuracy. Removing SimCLR or JEPA also degrades performance on several tasks, although JEPA removal produces negligible changes or slight improvements on RFS, LoS/NLoS, and Jamming. These effects indicate task-dependent contributions: removing any encoder degrades performance on at least three of the six tasks, providing evidence that all three objectives contribute useful features to the fitted probes.

\subsection{Subspace Analysis of Complementary Information}
\label{sec:residual}

Sections~\ref{sec:cka} and~\ref{sec:objective_importance} provide evidence of representational differences and useful contributions from all three objectives. We now examine how these contributions relate to MAE's representation subspace. For each encoder added to MAE (SimCLR or JEPA), we decompose its embeddings into a parallel component obtained by projection onto MAE's subspace and an orthogonal residual. We concatenate each component separately with MAE's embeddings and compare three conditions: \emph{base + full} combines MAE with the added encoder's complete embeddings; \emph{base + residual} combines MAE with only their orthogonal residual; and \emph{base + parallel} combines MAE with only their projection onto MAE's representation subspace.

The residual component retains reatained nearly all of the performance gain obtained from the full embeddings. For SimCLR, it recovers essentially all of the gain on positioning, beam prediction, and interference classification, but roughly half on RFS. For JEPA, it similarly retains the positioning and interference gains and most of the beam prediction gain. The parallel component, by contrast, adds effectively nothing: its gains are within noise of zero on every task, and in several cases performance drops slightly below the MAE-only baseline. These results suggest that much of the benefit of adding JEPA or SimCLR comes from components outside MAE's representation subspace, supporting the value of combining objectives that preserve complementary information.

\section{Conclusion and Future Work}
\label{sec:conclusion}

In this work, we developed a framework for scaling wireless foundation models through objective diversity. We introduced a fusion framework that combines frozen representations from independently pretrained encoders spanning reconstructive, predictive, and contrastive learning. To realize this diversity within the parameter budget of a single encoder, we proposed a multi-branch architecture that jointly trains a shared trunk and objective-specific branches on multiple self-supervised objectives.

Across six downstream tasks spanning spectrogram and CSI modalities, we showed that fusion outperforms the evaluated single-objective encoders at equal output embedding dimension while using roughly one-third of their parameters. In contrast, increasing single-objective encoder size produced inconsistent and sometimes negative returns. Our multi-branch architecture retained much of the fusion benefit within the parameter budget of a standard encoder. Through representation analyses, we found evidence of complementary contributions across objectives, with much of the benefit of adding predictive or contrastive representations to MAE retained in components orthogonal to its representation subspace. These findings support objective diversity as an effective strategy for allocating model capacity when wireless pretraining data are limited.

Future directions include evaluating larger pretraining corpora to examine how the benefits of objective diversity evolve with data scale, exploring adaptive loss weighting and shared-branch partitioning, and incorporating additional objectives and signal modalities to broaden the information captured by wireless foundation models.

\bibliography{bibliography}
\bibliographystyle{ieeetr}

\end{document}